\documentstyle[aps,twocolumn,prl,psfig]{revtex} 

\begin{document}

\draft
\twocolumn[    
\hsize\textwidth\columnwidth\hsize\csname @twocolumnfalse\endcsname    

\title{Competition between phase separation and ``classical'' intermediate
valence in an exactly solved model }
\author{Woonki Chung and J. K. Freericks}
\address{Department of Physics, Georgetown University, 
  Washington, DC 20057-0995}
\date{\today}
\maketitle

\widetext
\begin{abstract}
  The exact solution of the spin-$\frac{1}{2}$
  Falicov-Kimball model on an
  infinite-coordination Bethe lattice is analyzed in the regime
  of ``classical'' intermediate valence. We find
  (i) either phase separation  or a direct metal-insulator transition
  preclude intermediate valence over a large portion
  of the phase diagram and (ii)
  within the intermediate valence phase, only continuous
  transitions are found as functions of the localized f-electron
  energy or temperature.
\end{abstract}

\pacs{71.28+d, 71.30+h, and 71.10-Hf}
]      

\narrowtext
\paragraph*{Introduction}
The phenomenon of
intermediate valence (IV) is seen in a number of rare-earth compounds
(such as Ce, SmB$_6$, SmS, YbB$_{12}$, Eu$_3$O$_4$, etc.), where the average
f-electron filling per ion becomes nonintegral\cite{iv}. 
These materials
exhibit distinctive anomalies in thermodynamic and transport
properties accompanied with either continuous or discontinuous valence
transitions as the temperature, pressure, or composition are varied.

IV materials have a localized f-level that lies near the chemical potential
for the conduction electrons and is broadened by either hybridization
with the conduction band or by electron correlations.  This large density
of states acts as an electron resevoir and leads to both the IV phenomenon
and the thermal and transport anomalies.  Theoretical descriptions have
focused on two different approaches: (i) the Falicov-Kimball model (FKM)
\cite{FKM}, where the IV arises from an ensemble average of states with 
different
integral valence (``classical'' IV) and the transitions are either
discontinuous or continuous, being driven by the strength of the Coulomb
interaction between the f-electrons and conduction electrons, or (ii)
the periodic Anderson model (PAM)\cite{PAM}, where IV arises from a
quantum-mechanical mixture of states with different f-occupancy 
(``quantum-mechanical'' IV) and the transitions are continuous, being driven by
the hybridization between the f-electrons and the conduction electrons.  Most
materials fall into category (ii) and it is believed that the experimentally
observed discontinuous transitions arise from an additional coupling of the 
electrons to the lattice vibrations.  A realistic theoretical description 
should include the physics of both of these approaches, but has proven to
be cumbersome to carry out.  Here we focus on case (i) to see why few
materials can be found that fit into this scenario (likely materials 
\cite{iv,eu3s4} include
Eu$_3$O$_4$, Eu$_3$S$_4$, and Sm$_3$S$_4$).  We discovered that
the ``classical'' IV phase is often precluded by either phase separation or
a direct metal-insulator transition, which
we believe helps explain why few ``classical'' IV systems can be found in
nature.

It is important to examine the experimental differences between scenario (i)
and (ii).  The most obvious difference is in the response to a magnetic
field.  The FKM always possesses full f-moments, so it displays a Curie-like
susceptibility proportional to $1/T$.  If one includes the additional
(superexchange and RKKY) interactions between the moments, then the
uniform susceptibility will behave like $1/(T+T^*)$, which will not diverge 
for positive $T^*$.  Such a system will likely order in a spin-density wave
at some characteristic temperature, though.  The susceptibility never saturates
or displays a maximum as $T$ is lowered in the IV phase of the
FKM (it does in the metallic phase when
f-electrons are not present at low $T$ and become thermally populated at 
higher $T$ \cite{freericks_zlatic}).  The PAM, on the other 
hand, displays Curie-like behavior at high $T$ but either saturates, or
displays a low-$T$ maximum because of the Kondo effect, and the screening
of the local moments by the conduction electrons.  RKKY interactions are
also present and can lead to magnetic order, which makes a sharp differentiation
between the two models more difficult.  Most real materials that do not have
long-range magnetic order, display a susceptibility that either saturates or has
a low-$T$ maximum, indicating that scenario (ii) applies.  It is the purpose
of this contribution to explain why scenario (i) is so difficult to attain.

\paragraph*{Model}
The spin-$\frac{1}{2}$ FKM consists of localized f-electronic states and a
delocalized conduction band.  There is an on-site Coulomb interaction ($U>0$)
between the localized f-electron and the conduction electron.  The model
neglects the hybridization of the localized f-states with the
conduction band, and the valence transitions occur only when the
thermodynamic occupation of the different electronic states changes
under the variations of external conditions (such as pressure,
temperature, etc.).  The Hamiltonian of the model is
\begin{eqnarray}
  H &=& -\sum_{ij,\sigma}t_{ij}d^\dagger_{i\sigma}d_{j\sigma}
        +E_f\sum_{i,\sigma}f^\dagger_{i\sigma}f_{i\sigma} 
        +U_{ff}\sum_if^\dagger_{i\uparrow}f_{i\uparrow}
          f^\dagger_{i\downarrow}f_{i\downarrow} \nonumber \\
    & & \mbox{}+U\sum_{i,\sigma\sigma^\prime}d^\dagger_{i\sigma}d_{i\sigma}
          f^\dagger_{i\sigma^\prime}f_{i\sigma^\prime}    
        -\mu\sum_{i,\sigma}(d^\dagger_{i\sigma}d_{i\sigma} 
          +f^\dagger_{i\sigma}f_{i\sigma}) \,,
  \label{eq_H_particle}
\end{eqnarray}
where $d^\dagger_{i\sigma}$ ($d_{i\sigma}$) is the creation
(annihilation) operator for a conduction-band electron of spin
$\sigma$ at site $i$, $t_{ij}$ is the hopping matrix between lattice
sites $i$ and $j$, $f^\dagger_{i\sigma}$ ($f_{i\sigma}$) is the
creation (annihilation) operator for a localized electron with its
site energy $E_f$, and $U_{ff}$ is the on-site Coulomb repulsion
between f-electrons.  $U_{ff}$ is large in real materials,
so we choose $U_{ff} \rightarrow \infty$, and
restrict the number of f-electrons per site to $n_f \leq 1$. 
We choose the total number of electrons to satisfy $n_{total}=n_d+n_f=1$
to examine the IV phenomenon where each ion donates one electron to the system.
When the f-level lies below the bottom of the
conduction band, the system is an insulator ($n_d=0$, $n_f=1$); 
when the f-level lies above the middle of the conduction band, it becomes
a metal ($n_d=1$, $n_f=0$); IV phenomena can only occur when the f-level lies
inside the bottom half of the conduction band.  In our calculations,
we adjust a chemical potential $\mu$ to satisfy the constraint
$n_{total}=1$; $\mu$ is pinned to $E_f$ in the noninteracting IV regime, and the
average f-filling becomes nonintegral.

\paragraph*{Methodology}
The FKM  can be solved in the infinite-coordination-limit,
where the local approximation becomes exact, and the momentum
independent irreducible self-energy $\Sigma(\omega)$ has a
functional form which explicitly depends on $n_f$, $U$, and the local
Green's function $G(\omega)$.\cite{brandt,freericks_zlatic,mit_fkm}
We examine this model on the Bethe lattice, 
where the density of states for the noninteracting system becomes 
semicircular with the band width $4t^*=4t\sqrt{Z}$. [We take $t^*$ as our energy
unit ($t^*=1$).]  
In this case, there is a cubic equation for
$G(\omega)$ that determines the interacting density of
states $A(\omega)=-\frac{1}{\pi}{\rm Im}G(\omega)$ for any given
$n_f$\cite{van_Dongen}. Therefore, when $n_{total}=1$, we solve
the problem by minimizing the free energy\cite{FKM} $F[n_f,n_{total}=1]$ as a 
function of $n_f$ ($0\le n_f\le 1$):
\begin{eqnarray}
  F&[&n_f,n_{total}=1] = 2\int d\epsilon\,\epsilon\,f(\epsilon)\,A(\epsilon)
             + E_f\,n_f \nonumber \\
         & & \mbox{}+2T\int d\epsilon\,\left\{f(\epsilon) \ln f(\epsilon)
               +\left[1-f(\epsilon)\right]\ln[1-f(\epsilon)]
               \right\}A(\epsilon) \nonumber \\
         & & \mbox{}+T\left[n_f\ln n_f+(1-n_f)\ln(1-n_f)-n_f\ln2\right] \,,
  \label{eq_F}
\end{eqnarray}
where $f(\epsilon)$ is the Fermi distribution function. 

\begin{figure}
  \centerline{\psfig{figure=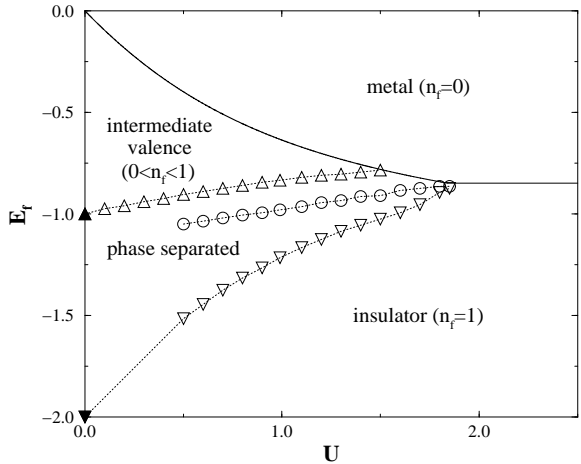,width=3.0in}}
  \caption{Phase diagram of the spin-$\frac{1}{2}$ Falicov-Kimball
    model when $n_{total}=1$ and $T=0$. 
    The filled-symbols and the solid lines indicate analytic results and
    the dotted lines are fit to the numerical results
    (open-symbols). The open circles denote the crossover from maximal
phase separation (below the circles) to IV phase separation (above the 
circles).}
  \label{f_phase}
\end{figure}

First, we construct the ground-state phase
diagram as a function of $E_f$ and $U$ (see Fig.~\ref{f_phase}).
For large enough $U$ \cite{mit_fkm}, there are only two phases: a metal when
$E_f>-\frac{8}{3\pi}$ and an insulator when $E_f<-\frac{8}{3\pi}$.  In
this limit, the system becomes effectively
noninteracting, avoiding the energetically unfavorable
double-occupancy of both d and f electrons on the same site, and does not
display IV.   For $U$ small enough, there is a range of values of $E_f$, lying
in the lower half of the conduction band, where the chemical potential is
pinned at $E_f$, and the average f-filling is nonintegral. As the
system changes from a metal to a homogeneous IV phase,
the value of the $n_f$, at which the ground-state energy
$F_{gs}[n_f,n_{total}=1]$ has its minimum, increases continuously from 0.
Hence, the boundary between the metal and the
homogeneous IV phases may be obtained from the
following condition:
\begin{equation}
  {\left.\frac{\partial F_{gs}}{\partial n_f}\right|}_{n_f=0,n_{total}=1}=\,0\ ,
  \label{eq_mi_cond}
\end{equation}
which determines when the metallic phase is no longer a local minimum of the
free energy.
The analytic form of $F_{gs}[n_f\!\rightarrow\!0, n_{total}=1]$ is found from
a low-density expansion\cite{mit_fkm} for
$\Sigma(\omega)$ and $G(\omega)$.  After some tedious
algebra, the boundary equation resulting from
Eq.~(\ref{eq_mi_cond}) becomes
\begin{eqnarray}
  E_f &=& -\frac{2}{\pi}+\frac{1}{2U}-\frac{U}{2}
          +\frac{2}{\pi}\left(\frac{1}{U}+U\right) \nonumber \\
      & & \mbox{}\times\left(\arctan\frac{2U}{1-U^2}
             -\arctan\frac{1+U}{1-U}\right)\,.
  \label{eq_mi}
\end{eqnarray}
Direct numerical calculations, minimizing the free energy in Eq.~(\ref{eq_F}) 
shows good agreement with
Eq.~(\ref{eq_mi}).  This result is valid for $U\le 1.84177$; larger values of
$U$ have the
direct transition from the metal to the insulator at $E_f=-8/3\pi$, 
which occurs before the 
metal becomes locally unstable.  

A similar analysis cannot be performed when 
$n_f\rightarrow 1$, because there is a first-order
transition between the insulator and a phase-separated state.
In order to show this phase separation, 
we minimize the free energy $F[n_f,n_{total}]$ 
with respect to $n_f$ for fixed $n_{total}$ and then determine the free-energy
curve as a function of $n_{total}$.  A Maxwell construction is finally
performed to
determine the convex hull of $F$ and see whether or not the unit-density
case is phase separated.  In equations, we compare
\begin{equation}
  F_{avg}=\alpha F[n^A_{total}]+(1-\alpha)F[n^B_{total}]\,,
  \label{eq_Favg}
\end{equation}
with
\begin{equation}
  1=\alpha n^A_{total}+(1-\alpha)n^B_{total}\,,
  \label{eq_alpha}
\end{equation}
to $F[n_{total}=1]$,
where the superscript $A$ indicates $n^A_{total}<1$ and $B$ indicates
$n^B_{total}>1$.

\paragraph*{Valence Transitions}
We begin at the $U=0$ limit.  In this case, IV phases can be found whenever 
$-2\le E_f\le 0$ and the chemical potential for the conduction
electrons is pinned at $E_f$ yielding $n_d$ conduction electrons 
$(0\le n_d\le 1)$.  The remaining electrons $1-n_d$ are f-electrons, and the
average filling per ion will be noninteger. However, because all of the 
f-electrons share the same energy, the ground-state energy of this configuration
is degenerate with any phase-separated mixture of states with different 
f-electron fillings (such as the integer-valent states
$n_f^A=0$, and $n_{total}^A=n_d$ and
$n_f^B=1$, and $n_{total}^B=n_d+1$) because the f-electron energy is
linear in the f-electron filling.  In order to determine what situation is
favorable as $U$ increases from zero, we need to expand the ground-state
energy in a power series through second order in $U$, and 
determine whether the homogenous phase, or the phase-separated states are
lower in energy.  Such an analysis is tedious, but shows that the 
maximal phase-separated state (where $n_f^A=0$ and $n_f^B=1$) is stable whenever
$E_f\le -1$. The rest of the phase diagram is determined
numerically in Fig.~1.  There are four 
different stable phases as a function of $E_f$ and $U$:  (i) the 
insulating phase, where all electrons are in the f-level, $n_d=0$ and
$n_f=1$; (ii) the
metallic phase, where all the electrons are in the conduction band, $n_d=1$ and 
$n_f=0$; (iii) the phase-separated state, which is maximal when
the A and B phases
that the system separates into have $n_f^A=0$ and $n_f^B=1$ respectively,
and is intermediate-valent when at least one 
state has nonintegral f-filling; and (iv) the homogeneous 
intermediate valence state, where $n_f$ is nonintegral.  Notice how both
phase separation and the direct metal-insulator transition
preclude IV behavior over much of the phase diagram.  The
phase separation is likely to become inhomogeneous charge ordering
in a real material, because the long-range Coulomb interaction (ignored in the
FKM) will not allow
the system to separate into states that have excess charge over a large volume,
and it will break up into microscopic domains of the different phases.

We examine this behavior in more detail in Figs.~2 and 3.  Fig.~2 shows the
Maxwell construction for the free energy in a case where phase separation
occurs.  Notice how the free energy becomes concave near $n_{total}=1$
which illustrates how the phase-separated state is stabilized.  Fig.~3 is
a vertical slice through the phase diagram at $U=0.9$.  It displays the
characteristics of all different phases.  When $E_f\le -1.265$ the
system is an insulator; for $-1.265\le E_f\le -0.995$ it is a maximal
phase separated state; for $-0.995\le E_f\le -0.845$ it is IV phase
separated, for $-0.845\le E_f\le -0.598$ it is in the homogeneous IV phase;
and for $-0.598\le E_f$ it is a metal.  There is no phase transition between 
the maximal and IV phase-separated states, rather it is a smooth crossover, 
occurring approximately at the position of the circles in Fig.~1 (our criterion
for the crossover is when $n_f^B\le 0.999$). 
We found that within the homogeneous 
IV phase, the valence change as a function of $E_f$ (and of $T$) was always
continuous, and exhibited no first-order (discontinuous) transitions.  The
only first-order transitions occur in the phase-separated states. Hence, the
only way to have discontinuous IV transitions arises either from the coupling
of the electrons to the lattice, or involves a phase-separation transition
rather than a pure IV change. (We should emphasize here that we have not
examined all other fillings of the FKM, where some discontinuous transitions
could occur.)  These results differ from those recently seen in the 
one-dimensional model \cite{farkasovsky}, 
where discontinuous IV transitions are also possible
(although that calculation has difficulty differentiating from a
discontinuous transition and phase separation).

These results summarize the behavior at $T=0$.  At finite temperatures, 
all of these phases survive, and phase transitions occur between them as
a function of temperature.  This is, in fact, the best place to look for these
first-order transitions experimentally---by measuring the specific heat as
a function of $T$ and looking for the large spikes near the first-order
transitions.  $C_V$ is determined by numerically differentiating the entropy (including averaging for the phase-separated states).
Typical results are summarized in Fig.~4, for all different ground-state phases.
All of our theoretical results are displayed in the experimental results of 
Eu$_3$S$_4$ \cite{eu3s4}: there is a first-order transition at 160K from a 
homogeneous to an inhomogeneous IV phase (with a sharp peak in the specific heat
accompanied by a structural transition) followed by a ferromagnetic transition
below 3.7~K (which is expected in any classical IV system which has local 
moments).

\paragraph*{Conclusions}

We have examined the phenomenon of intermediate valence in the spin-one-half
Falicov-Kimball model.  This is a model that can only display ``classical'' IV
phenomena, as the microscopic occupation of the f-electrons is always exactly
zero or one, but IV can occur from ensemble averaging, where the average
f-filling becomes nonintegral. Such systems should display magnetic 
susceptibilities that are Curie-like (or Curie-Weiss-like if additional magnetic
couplings are added in) which is not what is seen in most IV compounds.  Instead
most experiments show a susceptibility that either saturates, or has a maximum
as $T$ is lowered, which can best be described by the Kondo effect, and models
that include the hybridization between f-electrons and conduction electrons.
We discovered a fundamental reason why ``classical'' IV materials are more
difficult to find experimentally: either phase separation or a direct 
metal-insulator transition preclude the IV state
over a wide range of parameter space.  We believe this result helps explain
why nearly all observed IV materials require hybridization to describe them.
We also found, that the FKM does not support discontinuous IV transitions.
All valence changes, within the IV phase, are continuous as functions of 
$E_f$ or $T$.  The only ways to get discontinuous transitions IV is to either
couple the electrons to the lattice, or to have a transition to a 
phase-separated state \cite{disc}.

\section*{Acknowledgments}
We would like to acknowledge stimulating discussions with P. van Dongen and
V. Zlati\'c.
This work was supported by the Office of Naval Research Young
Investigator Program under the grant ONR N000149610828.



\begin{figure}[htbp]
  \centerline{\psfig{figure=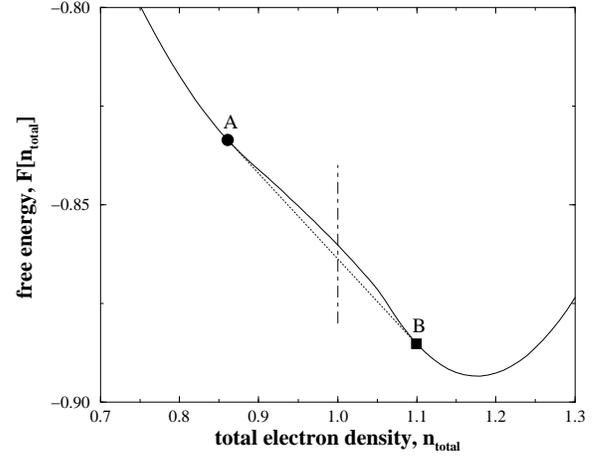,width=3.0in}}
  \caption{Maxwell construction of the free energy (solid line) as a function of
    the total electron density, which shows that the system phase
    separates into an A-phase ($n_{total}^A<1$, circle)
    and a B-phase ($n_{total}^B>1$, square) rather than remaining
    homogeneous with $n_{total}=1$.  The dotted
    line that connects A to B is the convex hull,
    and the vertical dot-dashed line is a
    guide to the eye for $n_{total}=1$.  Here, $U=1.5,\,E_f=-0.85$ and
    $T=0$.}
  \label{f_F}
\end{figure}

\begin{figure}[htbp]
  \centerline{\psfig{figure=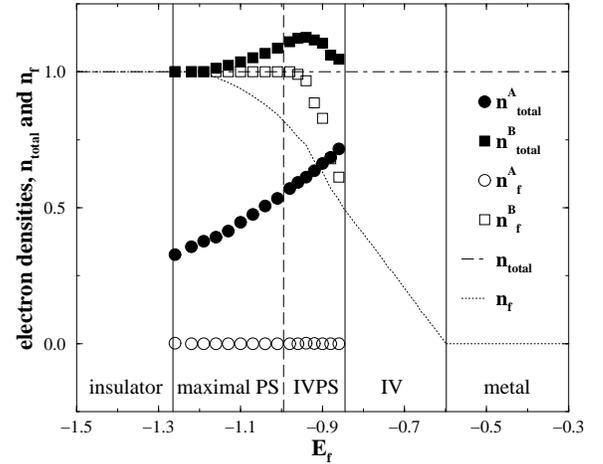,width=3.0in}}
  \caption{Electron density for all four ground-state phases at $U=0.9$ in the
    ground-state phase diagram.  The solid symbols show $n_{total}$
    and the open symbols denote $n_f$ for the phase-separated cases.
    The dot-dashed line is
    the average electron density, and the dotted line is the average f-electron
    density.  The system is an insulator for $E_f<-1.265$, phase separates when 
    $-1.265<E_f<-0.845$.  It has IV when $-0.995<E_f<-0.598$ and becomes a 
    simple metal for $E_f>-0.598$. The dashed line between the maximal and 
    IV phase-separated states marks the approximate location of the smooth 
    crossover.
    }
  \label{f_n}
\end{figure}

\begin{figure}
  \centerline{\psfig{figure=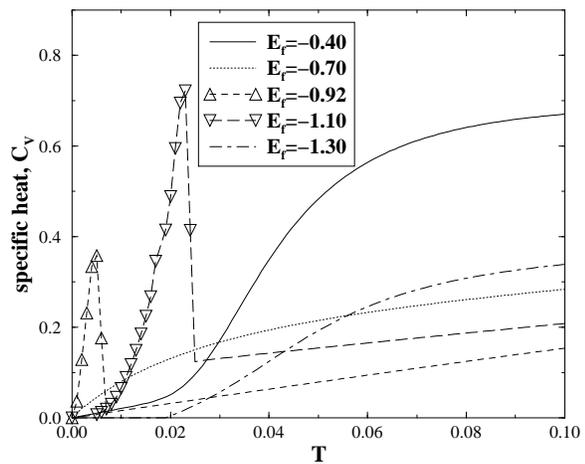,width=3.0in}}
  \caption{Specific heat for the various different values of $E_f$
    when $U=0.9$.  The sharp jump as the temperature decreases
    indicates the first-order transition (the phase separation)
    at the corresponding critical temperature. The triangles
    mark phase-separated states. }
  \label{f_Cv}
\end{figure}


\begin{references}
\bibitem{disc} We have not ruled out discontinuous IV
transitions at other values of $n_{total}$. The FKM does have discontinuous
metal-insulator transitions at larger values of U \cite{mit_fkm}.
\end{references}
\end{document}